\documentclass{aa}
\newcommand{\EQ}{\begin{equation}}
\newcommand{\EN}{\end{equation}}
\newcommand{\EQA}{\begin{eqnarray}}
\newcommand{\ENA}{\end{eqnarray}}

\newcommand{\Eq}[1]{Eq.~(\ref{#1})}
\newcommand{\Eqs}[2]{Eqs~(\ref{#1}) and~(\ref{#2})}

\newcommand{\Sec}[1]{Sect.~\ref{#1}}
\newcommand{\Fig}[1]{Fig.~\ref{#1}}
\newcommand{\FFig}[1]{Figure~\ref{#1}}
\newcommand{\Tab}[1]{Table~\ref{#1}}

\newcommand{\bra}[1]{\langle #1\rangle}

{}
{}
\newcommand{\meanuu}{\overline{\mbox{\boldmath $u$}}{}}{}
\newcommand{\meanoo}{\overline{\mbox{\boldmath $\omega$}}{}}{}
{}
{}
{}
{}
{}
\newcommand{\zz}{\hat{\mbox{\boldmath $z$}} {}}

\newcommand{\uu}{\mbox{\boldmath $u$} {}}

\newcommand{\BB}{\mbox{\boldmath $B$} {}}

\newcommand{\AAA}{\vec{A}}

\newcommand{\JJ}{\mbox{\boldmath $J$} {}}

\newcommand{\ff}{\mbox{\boldmath $f$} {}}

\newcommand{\nab}{\mbox{\boldmath $\nabla$} {}}

\newcommand{\oo}{\mbox{\boldmath $\omega$} {}}

\newcommand{\DD}{{\rm D} \, {}}

\newcommand{\ea}{{\rm et al. }}

\def\half{{\textstyle{1\over2}}}

\def\onethird{{\textstyle{1\over3}}}

\def\quarter{{\textstyle{1\over4}}}
\newcommand{\yapj}[3]{ #1, {ApJ,} {#2}, #3}
\newcommand{\yapjl}[3]{ #1, {ApJ,} {#2}, #3}

\newcommand{\yan}[3]{ #1, {Astr. Nachr.,} {#2}, #3}
\newcommand{\yana}[3]{ #1, {A\&A,} {#2}, #3}

\newcommand{\yjfm}[3]{ #1, {JFM,} {#2}, #3}
\newcommand{\ypf}[3]{ #1, {Phys. Fluids,} {#2}, #3}

\newcommand{\yjetp}[3]{ #1, {Sov. Phys. JETP,} {#2}, #3}
\newcommand{\yphy}[3]{ #1, {Physica,} {#2}, #3}

\newcommand{\yprs}[3]{ #1, {Proc. Roy. Soc. Lond.,} {#2}, #3}

\newcommand{\ymn}[3]{ #1, {MNRAS,} {#2}, #3}
\newcommand{\ynat}[3]{ #1, {Nat,} {#2}, #3}

\newcommand{\yjour}[4]{ #1, {#2}, {#3}, #4}
\newcommand{\ybook}[3]{ #1, {#2} (#3)}
\newcommand{\yproc}[5]{ #1, in {#3}, ed. #4 (#5), #2}

\begin{document}

\input{epsf}
\title{Astrophysical significance of the anisotropic kinetic alpha effect}
\author{A. Brandenburg\thanks{Also at: NORDITA, Blegdamsvej 17,
DK-2100 Copenhagen \O, Denmark} and B. v. Rekowski
}
\institute{
Department of Mathematics, University of Newcastle upon Tyne, NE1 7RU, UK}

\date{\today}

\abstract{ The generation of large scale flows by the anisotropic kinetic alpha
(AKA) effect is investigated in simulations with a suitable time-dependent
space- and time-periodic anisotropic forcing
lacking parity invariance. The forcing pattern moves
relative to the fluid, which
leads to a breaking of the Galilean invariance
as required for the AKA
effect to exist. The AKA effect is found to produce a clear large scale flow
pattern when the Reynolds number, ${\cal R}$, is small as only a few modes are
excited in linear theory.
In this case the non-vanishing components of the AKA tensor are dynamically independent
of the Reynolds number.
For larger values of ${\cal R}$, many more modes are excited and the components of the
AKA tensor are found to decrease rapidly with increasing value of ${\cal R}$.
However, once there is a magnetic field
(imposed and of sufficient strength, or dynamo-generated and saturated) the
field begins to suppress the AKA effect, regardless of the value of ${\cal R}$.
It is argued that the AKA effect is unlikely to be astrophysically significant
unless the magnetic field is weak and ${\cal R}$ is small.  }

\maketitle

\section{Introduction}

Vorticity and magnetic fields display some important similarities. Both
satisfy formally similar equations: the vorticity and induction
equations, respectively. This analogy has been used extensively by
Batchelor (1950) in early work on hydromagnetic turbulence and dynamos. Indeed,
both vorticity and magnetic field vectors are found to be aligned with
each other in simulations of convection with dynamo-generated magnetic
fields (Brandenburg \ea 1996). However, the intensity of the two vector
fields does not generally show any correlation, and even slightly different
initial conditions for vorticity and magnetic field lead to their mutual
departure after some time (Moffatt 1978).

A formal analogy between vorticity and magnetic fields has also been proposed
in the context of mean-field theory. Moiseev \ea (1983) proposed the
possibility of an alpha effect in the equation for the mean vorticity in a
compressible fluid. However, if a similar effect is to exist in the
incompressible case there must be some 
anisotropic forcing, because
otherwise the non-dissipative terms in the equation for the mean velocity would
vanish (Krause \& R\"udiger 1974, see also Moffatt \& Tsinober 1992).

Frisch \ea (1987) were the first to study the effects of
a non-Galilean invariant
forcing that produced a destabilization of the velocity field at large
scales. In analogy with the alpha effect in dynamo theory (Moffatt 1978) they called
this the {\it anisotropic kinetic alpha effect}, or AKA effect. The
nonlinear behaviour of this effect was studied by Sulem \ea (1989) and
Galanti \& Sulem (1991), who
showed that the energy transfer to larger scales happens successively,
in the form of an inverse cascade.

In recent years a number of geophysical and astrophysical applications
of the AKA effect have been discussed in the literature. For example, Zimin \ea (1989),
Rutkevich (1993), and Levina \ea (2000) applied the effect to geophysical convection.
Krishan (1991) discussed the possibility of an inverse cascade in the
context of solar granulation and Krishan (1993) invoked the anisotropic
kinetic alpha effect in order to explain the clustering of galaxies.
Kitchatinov \ea (1994) and v.\ Rekowski \ea (1999) discussed the
possibility of creating large scale vortices in accretion discs. Recently,
v.\ Rekowski \& R\"udiger (1998) suggested that the AKA effect could help
to solve the `Taylor number puzzle' in models of stellar differential
rotation. They found that this effect could produce angular velocity
contours that are no longer constant on cylinders, and hence closer to
the helioseismological observations.

In magnetohydrodynamics the generation of large scale magnetic fields has been
well established numerically in simulations of the geodynamo (Glatzmaier \&
Roberts 1995) and accretion discs (Brandenburg \ea 1995), for example. However,
the excitation condition for the generation of large scale vortices should be
similar to the condition for dynamo action (Kitchatinov \ea 1994). It is
therefore remarkable that no evidence for the spontaneous generation of
vortices has been seen in the simulations of Brandenburg \ea (1995), even
though the boundary conditions for vorticity and magnetic field were identical.
The same is true of the recent simulations of Brandenburg (2001) using fully
helical isotropic flows in periodic domains. No trends for large scale flows
are seen, even though the magnetic field displays a very pronounced large scale
pattern.

The evidence for large scale flows is sparse. However, one example might be the
large scale flows seen in highly supercritical Rayleigh-Benard convection
(Howard \& Krishnamurti 1986). Here it could be the presence of boundaries
which breaks Galilean invariance. By contrast, the simulations of
Brandenburg (2001) were Galilean invariant and isotropic, which explains the absence of an
AKA effect. Astrophysical examples where turbulence is driven by
non-Galilean invariant forcing include supernova-driven turbulence
in galaxies (Korpi \ea 1999) and the turbulent wakes driven by
galaxies moving through the galaxy cluster (Ruzmaikin \ea 1989).

It is important to realize that the AKA effect has been verified numerically
only in the case of rather low Reynolds numbers, ${\cal R} \le 2$. Given the
possible astrophysical relevance of the AKA effect, it is important to assess
the dependence of the resulting large scale flow on the Reynolds number, which
is extremely high in all astrophysical settings.  Another important property of
astrophysical flows is that the gas is ionized and electrically conducting, so
it may be unstable to dynamo action. Typically, the resulting large scale
magnetic fields attain near-equipartition strength and will therefore be
dynamically important. The combined action of AKA and $\alpha$ effects has
already been considered by Galanti \ea (1990, 1991), but again only at
relatively small kinetic and magnetic Reynolds numbers.

The purpose of the present paper is to extend the studies of
Frisch \ea (1987) and Sulem \ea (1989) to the case of larger Reynolds
numbers, allowing also for a magnetic field to grow from a weak initial
seed magnetic field. In all cases we adopt the same forcing as
Frisch \ea (1987).
However, in contrast to their original paper, where the flow
was assumed to be incompressible, we assume here weak compressibility.

\section{Description of the model}

We solve the isothermal compressible hydromagnetic equations for the logarithmic
density $\ln\rho$, the velocity $\uu$, and the magnetic vector potential
$\AAA$,
\EQ
{\DD\ln\rho\over\DD t}=-\nab\cdot\uu,
\label{dlnrhodt}
\EN
\EQ
{\DD\uu\over\DD t}=-c_{\rm s}^2\nab\ln\rho+{\JJ\times\BB\over\rho}
+{\mu\over\rho}(\nabla^2\uu\!+\!\onethird\nab\nab\!\cdot\!\uu)+\ff,
\label{dudt}
\EN
\EQ
{\partial\AAA\over\partial t}=\uu\times\BB-\eta\mu_0\JJ,
\label{dAdt}
\EN
in a three-dimensional periodic cartesian domain of size $L = 2\pi/k_1$, where $k_1$ is the
smallest wavenumber in the box,
${\rm D}/{\rm D}t=\partial/\partial t+\uu\cdot\nab$ the
advective derivative, $\BB=\nab\times\AAA$ the magnetic flux density,
$\JJ=\nab\times\BB/\mu_0$ the current density, $c_{\rm s}$ the sound speed, and
\EQ
\ff=\sqrt{2}f_0\pmatrix{\phi_1\cr \phi_2\cr \phi_1+\phi_2}
\EN
is the forcing term of Frisch \ea (1987) with
\EQ
\phi_1=\cos(k_{\rm f}y+\omega t),\quad
\phi_2=\cos(k_{\rm f}x-\omega t),\quad
\EN
and $\omega=\nu k_{\rm f}^2$.
This forcing corresponds to a pattern
moving with the velocity $(1,-1,0)\nu k_{\rm f}$ diagonally in the $(x,y)$-plane.
$k_{\rm f}$ is the wavenumber of the small scale forcing and $f_0$
($=\mbox{const}$) gives the strength of the forcing.
The (uncurled) induction equation (\ref{dAdt}) implies a specific
gauge for $\AAA$ such that the electrostatic potential vanishes. Instead
of the dynamical viscosity $\mu$ ($=\mbox{const}$ -- not to be confused
with the magnetic permeability $\mu_0$) we will in the following refer to
the {\it mean} kinematic viscosity
$\nu\equiv\mu/\rho_0$, where $\rho_0$ is the volume averaged density
($\rho_0=\mbox{const}$ owing to mass conservation).
$\eta$ ($=\mbox{const}$) is the magnetic diffusivity.

In order to nondimensionalize the equations, velocity is measured in units of
$\nu k_{\rm f}$, length in units of $k_{\rm f}^{-1}$, time in units of
$(\nu k_{\rm f}^2)^{-1}$ ($\equiv\omega^{-1}$) and magnetic field in units of
$\sqrt{\mu_0\rho_0}\nu k_{\rm f}$.
The nondimensional form of \Eq{dlnrhodt} is then unchanged, and the
nondimensional momentum and induction equations become
\EQ
{\DD \uu \over \DD t} = -{{\cal R}^2\over {\cal M}^2}\nab \ln\rho
                             +{\JJ \times \BB \over \rho}
+{\nabla^2\uu\!+\!\onethird\nab\nab\!\cdot\!\uu\over\rho}+{\cal R}\ff,
\label{dudt_nondim}
\EN
\EQ
{\partial\AAA \over \partial t} =  \uu \times \BB
                                  -{1\over {\cal P}_{\rm m}}\JJ,
\label{dAdt_nondim}
\EN
where all variables are nondimensional, and the nondimensional forcing
function is $\ff=\sqrt{2}\;(\phi_1, \phi_2, \phi_1+\phi_2)$. In the
following, however, we express all relevant variables in explicitly
nondimensional form, e.g.\ we write $\omega t$ instead of just $t$.

The problem is completely defined by four nondimensional parameters: the
Reynolds number ${\cal R}=U_0/(\nu k_{\rm f})$, where $U_0\equiv f_0/\omega$ is
a reference velocity, the magnetic Reynolds number ${\cal R}_{\rm m}=U_0/(\eta
k_{\rm f})$, the Mach number ${\cal M}=U_0/c_{\rm s}$, and the size of the
computational domain $L$, which we express in terms of the nondimensional
quantity $k_{\rm f}/k_1=k_{\rm f}L/(2\pi)$. The magnetic Prandtl number is
${\cal P}_{\rm m}={\cal R}_{\rm m}/{\cal R}$. In some additional runs we also
applied an external field $\BB_0=B_0\zz$, which leads to a fifth parameter
${\cal B}_0\equiv B_0/(\sqrt{\mu_0\rho_0}U_0)$. We note that our definition of
${\cal R}$ agrees with that of Frisch \ea (1987).

As a representative measure of the resulting velocity field we monitor the
normalized rms velocity, ${\cal U}=\bra{\uu^2}^{1/2}/U_0$. Here the angular
brackets denote averaging over the volume of the domain. The magnetic field is
monitored analogously through ${\cal
B}=\bra{\BB^2}^{1/2}/(\sqrt{\mu_0\rho_0}U_0) + {\cal B}_0$, where ${\cal B}_0$
quantifies the imposed magnetic field. However, in most of the cases where we
have a magnetic field we rely on the dynamo-generated field and put ${\cal
B}_0=0$. We also calculate the normalized rms velocity of the large scale flow,
${\cal U}_{\rm LS}=\bra{\meanuu^2}^{1/2}/U_0$. With the forcing function chosen
the $z$-direction is preferred and horizontal $(x,y)$ averages (denoted by
overbars) are appropriate for extracting the large scale flow. We then give the
ratio ${\cal U}_{\rm LS}/{\cal U}$. We note, however, that a definition of
${\cal U}_{\rm LS}$ in terms of Fourier filtering to $k=k_1$ is also sensible.
In some cases we give the relative kinetic helicity
of the horizontally averaged (large scale) flow,
\EQ
{\cal H}_{\rm K}^{\rm LS}
=\bra{\meanoo\cdot\meanuu}/(\bra{\meanoo^2}\bra{\meanuu^2})^{1/2},
\EN
where angular brackets indicate volume averages (as opposed to the
overbars which denote only horizontal averages). We also
calculate the nondimensional growth rate of the magnetic field,
$\lambda/\omega$, the ratio of rms magnetic field to rms velocity, ${\cal
B}/{\cal U}$, as well as nondimensional measures of kinetic and magnetic
helicities, \EQ {\cal H}_{\rm
K}=\bra{\oo\cdot\uu}/(\bra{\oo^2}\bra{\uu^2})^{1/2}, \EN \EQ {\cal H}_{\rm
M}=\bra{\AAA\cdot\BB}/(\bra{\AAA_{\rm c}^2}\bra{\BB^2})^{1/2}, \EN
respectively. Here, $\AAA_{\rm c}=\AAA-\AAA_0$
is the magnetic vector potential in Coulomb gauge, where
$\AAA_0=\bra{\AAA}+\nab\phi$ and $\nabla^2\phi=\nab\cdot\AAA$.
The advantage of using $\AAA_{\rm c}$ is that it has the property of
minimizing $\bra{\AAA^2}$. The quantity $\bra{\AAA\cdot\BB}$ is gauge invariant
in periodic domains (and therefore $\bra{\AAA_{\rm c}\cdot\BB}=\bra{\AAA\cdot\BB}$,
for example). If there is an additional imposed uniform magnetic
field, $\bra{\AAA\cdot\BB}$ would no longer be gauge invariant,
because $\bra{\AAA}\neq0$ in general. Since
a uniform field in isolation has zero magnetic helicity we calculate
$\bra{\AAA\cdot\BB}$ only with respect to the deviations from the
imposed field.
Finally, we also monitor the quantity
\EQ
{\tilde{\cal H}}_{\rm M}=k_1\bra{\AAA\cdot\BB}/\bra{\BB^2},
\EN
which is a nondimensional length scale. The reason for using here $k_1$
instead of $k_{\rm f}$ is that the magnetic helicity tends to develop on 
the largest possible scale, independent of the forcing wavenumber.

\section{Results}

\begin{figure}[t!]\epsfbox{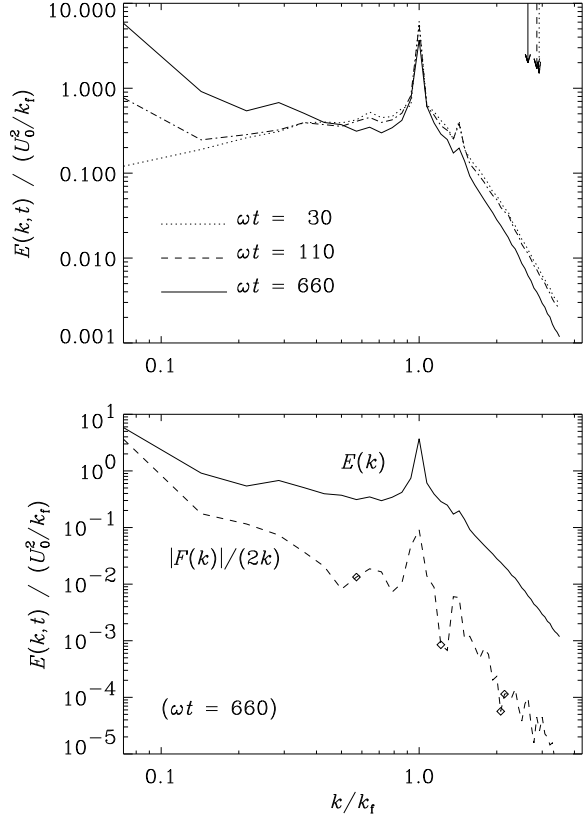}\caption[]{ Upper
panel: Power spectra of kinetic energy, $E(k,t)$, for Run~9c at three different
times. Note the gradual build-up of power in the lowest wavenumber, $k_1=k_{\rm
f}/14$. For reference, the dissipative cutoff wavenumber, defined as $k_{\rm
d}\equiv(\bra{\oo^2}/\nu^2)^{1/4}$, is indicated by arrows in the upper-right
corner separately for the three times. Lower panel: $E(k)$ (at $t=660/\omega$)
is compared with the helicity spectrum, $F(k)$, normalized by $2k$. The
diamonds mark the few points where $F(k)>0$; everywhere else $F(k)<0$. Note
that only at the smallest wavenumber the helicity is close to its largest
possible value ($|F|\leq 2kE$ by the realizability condition). ${\cal R}=8$.
}\label{Fpspec_Aka3}\end{figure}

\begin{figure}[h!]\epsfbox{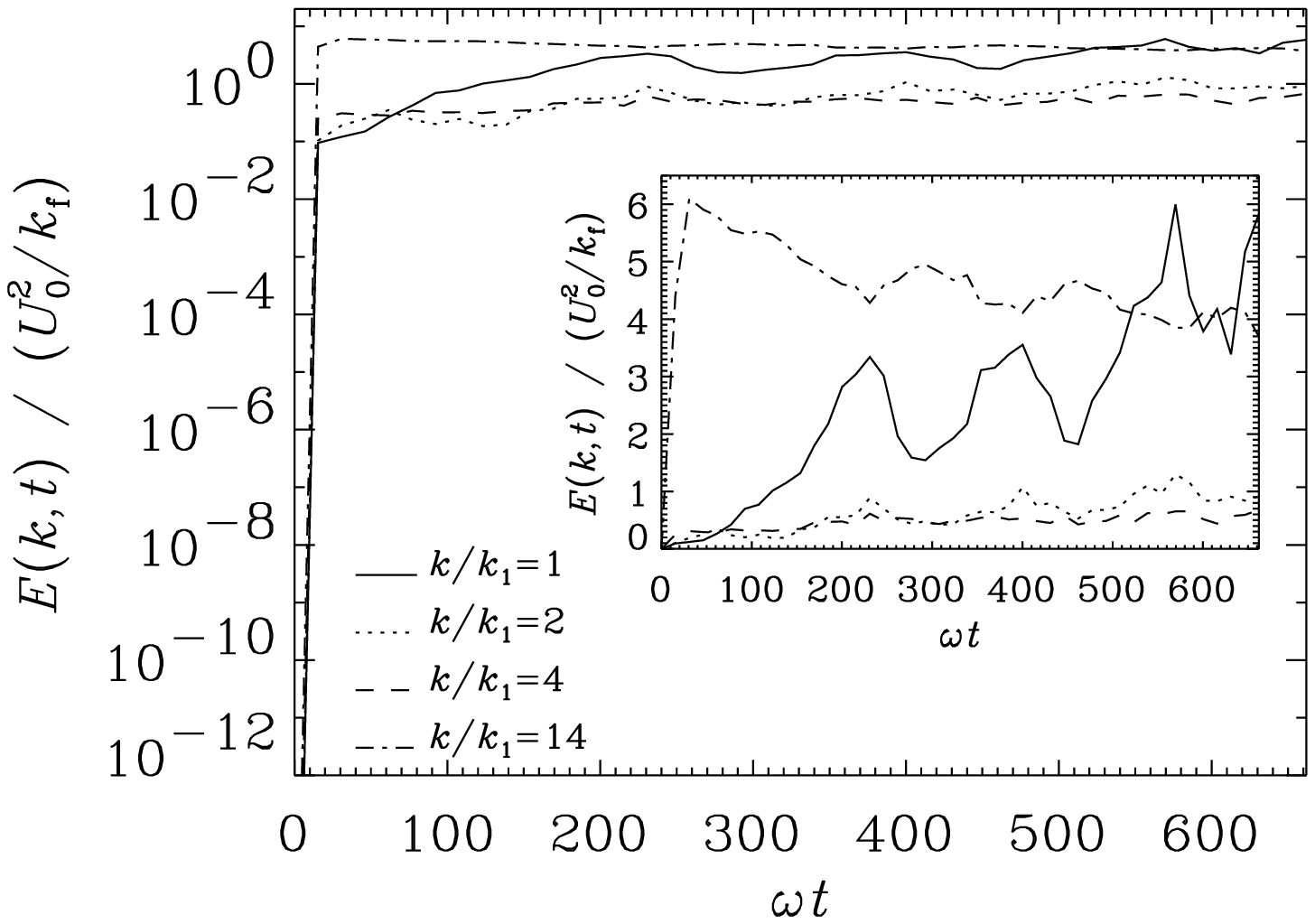}\caption[]{ Evolution
of spectral energy in different modes in a lin-log plot. The inset shows the
same, but here the ordinate scale is linear. Run~9c.
}\label{Fpekt_Aka3}\end{figure}

We first consider the case where ${\cal R}_{\rm m}$ is so small there is no
dynamo action. We have calculated solutions for different values of ${\cal R}$
(between $\sqrt{2/3}\approx0.82$ and 12) and $k_{\rm f}/k_1$ (between 6 and
14). The case ${\cal R}=\sqrt{2/3}$ and $k_{\rm f}/k_1=6$ was considered by
Frisch \ea (1987) for the strictly incompressible case and neglecting any
magnetic fields completely. Without magnetic fields of sufficient strength we
find in all cases clear signs of the AKA effect, which manifests itself through
enhanced power in the lowest Fourier modes of the kinetic energy spectrum; see
\Fig{Fpspec_Aka3}. It turns out that at least for small values of ${\cal R}$
the spectral power in the lowest Fourier mode exceeds that of the forcing
wavenumber. As the value of ${\cal R}$ is increased the solution becomes more
irregular in time (\Fig{Fpekt_Aka3}), as also found by Sulem \ea (1989). It is
also noteworthy that, at least for moderately large values of ${\cal R}$, the
spectral peak at the smallest wavenumber is relatively broad compared with the
magnetic inverse cascade where, again using a periodic box, the peak is very
sharp; cf.\ Figs.~17 and 19 of Brandenburg (2001).

\epsfxsize=8.2cm

\epsfxsize=8cm

\begin{table*}[t!]\caption{
Summary of the main properties of various runs. A hyphen in the entry for
${\cal B}$ indicates that the magnetic field decays.
}\vspace{12pt}\centerline{\begin{tabular}{cc|ccccc|ccccccc} Run & mesh &
$k_{\rm f}/k_1$ & ${\cal R}$ & ${\cal R}_{\rm m}$ & ${\cal M}$
    & ${\cal B}_0$
    & ${\cal U}$ & ${\cal U}_{\rm LS}/{\cal U}$
    & $\lambda/\omega$ & ${\cal B}/{\cal U}$
    & ${\cal H}_{\rm K}$ & ${\cal H}_{\rm M}$ & $\tilde{\cal H}_{\rm M}$
\\
\hline
1a & $32^3$ &  6      & 0.82 & 0.82    & 0.15 & 0         & 2.40    & 0.86    &-0.04     & --      & -0.19 & --        & --    \\ 
1b & $32^3$ &  6      & 0.82 & \bf{41} & 0.15 & 0         & 1.36    & 0.09    & 0.05     & 0.24    & -0.06 & 0.00      & 0.00  \\ 
1c & $32^3$ &  6      & 0.82 & 0.82    & 0.15 & \bf{0.07} & 2.10    & 0.78    & 0.02     & 0.24    & -0.16 & --        & --    \\ 
1d & $32^3$ &  6      & 0.82 & 0.82    & 0.15 & \bf{0.14} & 1.67    & 0.60    & 0.01     & 0.39    & -0.09 & --        & --    \\ 
\hline
2a & $30^3$ &  6      & 1    & 0.33    & 0.10 & 0         & 1.46    & 0.73    & --       & --      & -0.14 & --        & --    \\ 
\hline
3a & $30^3$ &  6      & 1.2  & 0.4     & 0.12 & 0         & 1.60    & 0.81    & --       & --      & -0.19 & --        & --    \\ 
\hline
4a & $30^3$ &  6      & 1.5  & 0.5     & 0.15 & 0         & 1.61    & 0.85    & --       & --      & -0.21 & --        & --    \\ 
4b & $30^3$ & \bf{9}  & 1.5  & 1.5     & 0.15 & 0         & 2.91    & 0.87    &-0.02     & --      & -0.26 & --        & --    \\ 
4c & $30^3$ &  9      & 1.5  & \bf{15} & 0.15 & 0         & 1.55    & 0.08    & 0.15     & 0.84    & -0.04 & 0.35      & 0.19  \\ 
\hline
5a & $30^3$ &  6      & 2    & 0.67    & 0.20 & 0         & 1.52    & 0.88    & --       & --      & -0.22 & --        & --    \\ 
5b & $60^3$ & \bf{14} & 2    & 2       & 0.18 & 0         & 1.77    & 0.44    &-0.01     & --      & -0.08 & --        & --    \\ 
5c & $60^3$ & 14      & 2    & \bf{20} & 0.18 & 0         & 1.29    & 0.04    & 0.20     & 0.83    & -0.05 & 0.29      & 0.11  \\ 
\hline
6a & $30^3$ &  6      & 2.3  & 0.77    & 0.23 & 0         & 1.48    & 0.89    & --       & --      & -0.23 & --        & --    \\ 
\hline
7a & $30^3$ &  6      & 3    & 1       & 0.30 & 0         & 1.43    & 0.84    & --       & --      & -0.23 & --        & --    \\ 
7b &\bf{$60^3$}&  6   & 3    & 1       & 0.06 & 0         & 1.78    & 0.77    & 0.00     & --      & -0.20 & --        & --    \\ 
7c & $60^3$ & \bf{9}  & 3    & 1       & 0.06 & 0         & 1.43    & 0.41    &-0.06     & --      & -0.10 & --        & --    \\ 
7d & $60^3$ & \bf{14} & 3    & 1       & 0.06 & 0         & 1.56    & 0.32    &-0.03     & --      & -0.09 & --        & --    \\ 
7e & $60^3$ & \bf{9}  & 3    & --      & 0.17 & 0         & 1.60    & 0.38    & --       & --      & -0.10 & --        & --    \\ 
7f & $60^3$ & \bf{14} & 3    & --      & 0.17 & 0         & 1.66    & 0.45    & --       & --      & -0.08 & --        & --    \\ 
7g & $60^3$ &  9      & 3    & \bf{12} & 0.17 & 0         & 1.12    & 0.12    & 0.02     & 0.90    & -0.09 & 0.39      & 0.21  \\ 
\hline
8a & $60^3$ &  6      & 5    & 5       & 0.25 & 0         & 1.22    & 0.44    &-0.07     & --      & -0.11 & --        & --    \\ 
8b & $60^3$ & \bf{9}  & 5    & 5       & 0.10 & 0         & 1.20    & 0.26    &-0.03     & --      & -0.11 & --        & --    \\ 
\hline
9a & $60^3$ &  6      & 8    & 2.7     & 0.27 & 0         & 1.00    & 0.29    &-0.19     & --      & -0.12 & --        & --    \\ 
9b & $80^3$ & \bf{9}  & 8    & 2.7     & 0.27 & 0         & 1.04    & 0.36    &-0.08     & --      & -0.09 & --        & --    \\ 
9c &$100^3$ & \bf{14} & 8    & 8       & 0.17 & 0         & 1.11    & 0.44    &-0.01     & --      & -0.06 & --        & --    \\ 
9d & $81^3$ & 14      & 8    & \bf{40} & 0.35 & 0         & 0.84    & 0.07    & 0.39     & 0.65    & -0.17 & 0.22      & 0.11  \\ 
\hline
10a & $60^3$ &  6      & 12  & 12      & 0.33 & 0         & 0.88    & 0.40    &-0.05     & --      & -0.09 & --        & --    \\ 
10b & $60^3$ & \bf{9}  & 12  & 12      & 0.33 & 0         & 0.90    & 0.33    &-0.04     & --      & -0.09 & --        & --    \\ 
10c & $80^3$ & \bf{12} & 12  & 12      & 0.17 & 0         & 0.90    & 0.31    &-0.04     & --      & -0.07 & --        & --    \\ 
10d & $60^3$ &  9      & 12  & \bf{60} & 0.33 & 0         & 0.78    & 0.06    & 0.72     & 0.36    & -0.23 & 0.03      & 0.004 \\ 
\label{T1}\end{tabular}}\end{table*}

A summary of all the runs that we have performed is given in \Tab{T1}. Here we
list some characteristic quantities as a function of ${\cal R}$ and other input
parameters. Perhaps the most important diagnostic quantity is the ratio ${\cal
U}_{\rm LS}/{\cal U}$ characterizing the relative importance of the resulting
large scale flow. As a rule, when ${\cal U}_{\rm LS}/{\cal U}>0.7$, the
spectral energy at the largest scale exceeds that at the forcing scale. For
${\cal U}_{\rm LS}/{\cal U}$ in the range 0.4 to 0.5 the two are comparable on
average, but the large scale flow is unsteady in time. For ${\cal U}_{\rm
LS}/{\cal U}$ less than about 0.3 there is usually only the one peak at the
forcing scale. This is the case especially when there is a magnetic field,
i.e.\ ${\cal B}/{\cal U}$ is finite.

In the next subsection we discuss hydrodynamic aspects, i.e. we assume that
${\cal R}_{\rm m}$ is below the critical value for dynamo action, which is
between 10 and 20. The discussion of runs with dynamo action follows in
\Sec{Smagnetic}.

\subsection{Nature of the large scale flow}

\begin{figure}[h!]\epsfbox{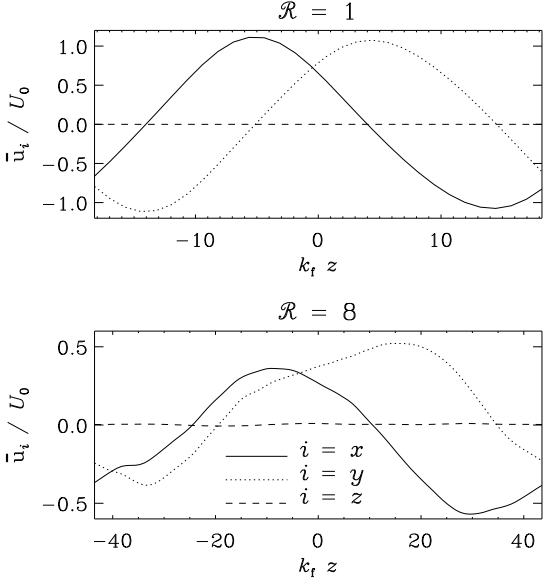}\caption[]{ The
three horizontally averaged velocity components for Run~2a (${\cal R}=1$,
$k_{\rm f}/k_1=6$) and Run~9c (${\cal R}=8$, $k_{\rm f}/k_1=14$). The relative kinetic
helicity of the depicted large scale field is ${\cal H}_{\rm K}^{\rm LS}=-0.99$
for Run~2a and $-0.79$ for Run~9c.}\label{Fpaver_comp}\end{figure}

In all cases the solutions have a slowly varying large scale component, clearly
seen in the horizontal averages (denoted by overbars) of the velocity.
\FFig{Fpaver_comp} shows that ${\overline u}_x(z)$ and ${\overline u}_y(z)$
vary approximately sinusoidally in $z$ and are phase shifted relative to each
other by about 90 degrees. This type of mean flow corresponds to a Beltrami
wave which is helical ($\meanoo\cdot\meanuu$ is here negative;
$\meanoo=\nab\times\meanuu$) and approximately force-free (i.e.\ it has
no inertia; $\meanuu\cdot\nab\meanuu=0$). Because of weak compressibility,
$\nab\cdot\meanuu$ is not strictly zero and therefore ${\overline u}_z$ does
not need to vanish exactly, but it is nevertheless very close to zero; see the
dashed lines in \Fig{Fpaver_comp}.

\epsfxsize=8cm

\begin{figure}[h!]\epsfbox{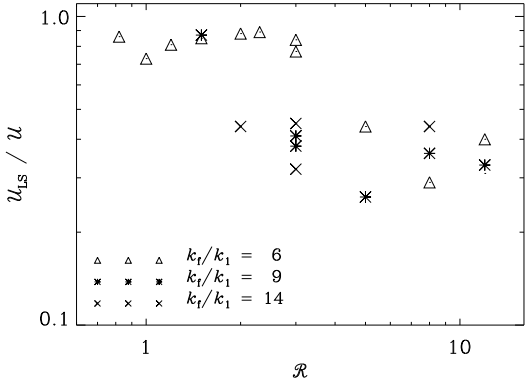}\caption[]{ The relative
importance of the large scale flow, as measured by ${\cal U}_{\rm LS}/{\cal
U}$, versus ${\cal R}$. Different values of $k_{\rm f}/k_1$ are indicated by
different symbols. Note the segregation into two different groups.
}\label{Fpuls}\end{figure}

The rms velocity of the large scale flow relative to that of the
full velocity field is given by ${\cal U}_{\rm LS}/{\cal U}$.
As the value of ${\cal R}$ is increased the relative importance of
the large scale flow diminishes; see \Fig{Fpuls}.
This gives us a first indication that in the astrophysically
interesting limit of ${\cal R}\rightarrow\infty$ the AKA-generated
large scale flow will be less prominent than in the case where
${\cal R}$ is small.

\epsfxsize=8cm

Next we analyse the relation between the mean (large scale) flow and components of
the Reynolds stress, $R_{ij}\equiv{\overline{u'_i u'_j}}$, where primes
denote deviations from the mean flow, i.e.\ $\uu'=\uu-\meanuu$. To
lowest order the AKA effect couples the components of the Reynolds
stress linearly to the mean flow,
\EQ
R_{ij}={\rm const}-\alpha_{ijk}{\overline u}_k+
\mbox{higher order terms},
\label{alpha_def}
\EN
which closes the equations for the mean flow.
Similarly to the alpha effect in dynamo theory, the $\alpha_{ijk}$ term
produces a linear instability for wavenumbers
$k/k_{\rm f}<\half{\cal R}^2$ (Frisch \ea 1987). The growth rate is
maximum when $k/k_{\rm f}=\quarter{\cal R}^2$. For example, when
${\cal R}^2=2/3$ (the value considered by Frisch \ea 1987) then
the fastest growth is for $k/k_{\rm f}=1/6$, which was also the
smallest wavenumber ratio in their model.
Moreover, in this case, this is also the only unstable mode;
$k/k_{\rm f}=1/3$ is already stable and there are no other discrete
modes in-between.

For small values of ${\cal R}$, Frisch \ea (1987) found that
when the large scale flow reaches
saturation, the Reynolds stress tensor depends nonlinearly on
$\meanuu$. In particular, its horizontal off-diagonal components
are given by
\EQ
R_{xz}={\half U_0^2\over1+{\overline u}_y/(\nu k_{\rm f})
+\half{\overline u}_y^2/(\nu k_{\rm f})^2},
\label{stressxz}
\EN
\EQ
R_{yz}={\half U_0^2\over1-{\overline u}_x/(\nu k_{\rm f})
+\half{\overline u}_x^2/(\nu k_{\rm f})^2}.
\label{stressyz}
\EN
For small values of $|\meanuu|$ one recovers \Eq{alpha_def} with
\EQ
\alpha_{xzy}=-\alpha_{yzx}
=\half U_0^2/(\nu k_{\rm f})
=\half{\cal R}U_0.
\EN

\begin{figure}[h!]\epsfbox{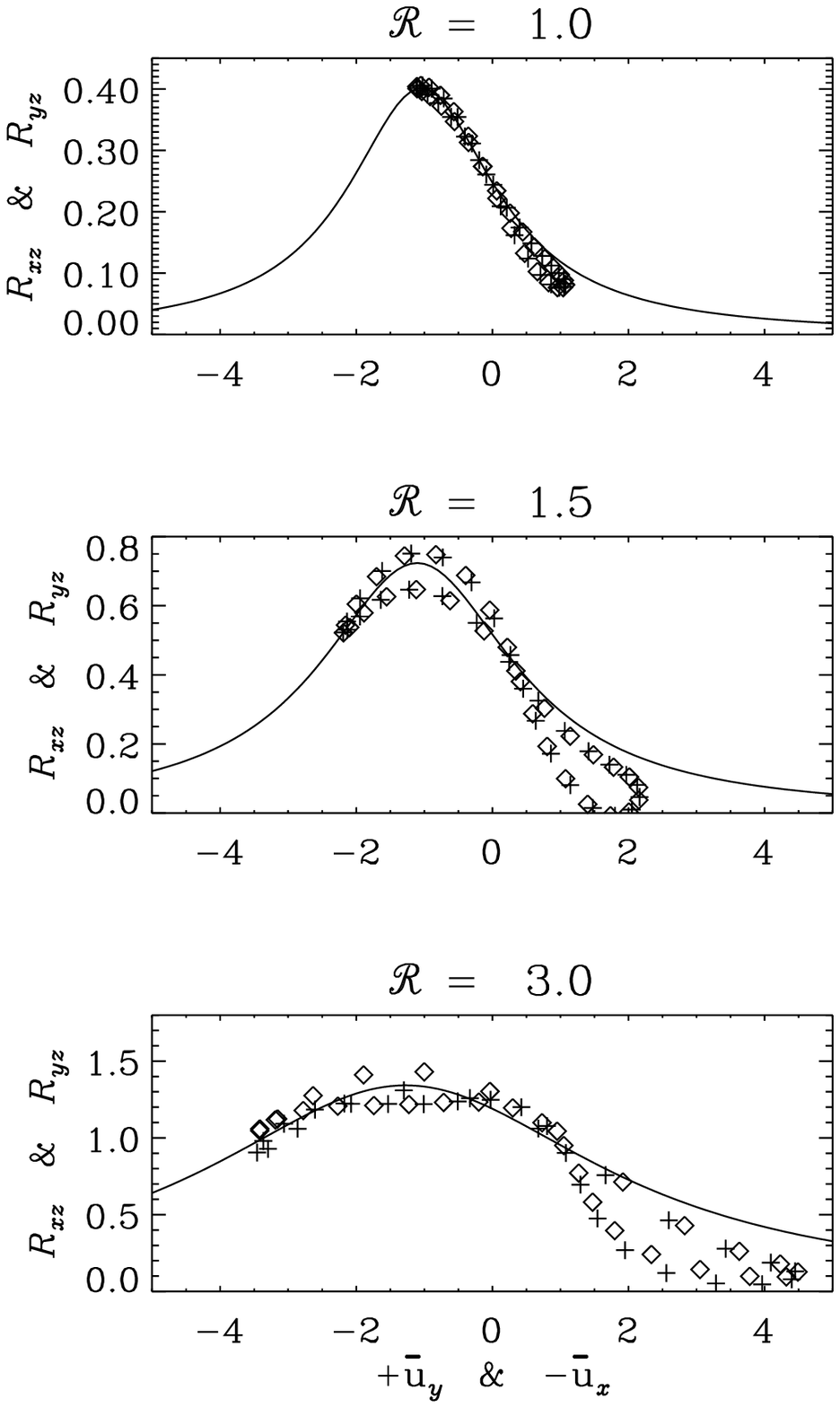}\caption[]{
$R_{xz}$ vs ${\overline u}_y$ (plus signs) superimposed with $R_{yz}$ vs
$-{\overline u}_x$ (diamonds).
The solid line gives the fit \Eq{stressfitxy} with the parameters $q_0$, $q_1$,
and $q_2$ from \Tab{Tfit}.}\label{Fpstress_avero_comp}\end{figure}

In order to check \Eqs{stressxz}{stressyz} for different values of ${\cal R}$
we generalize this relation to \EQ R_{xz}={q_0\over1+q_1{\overline
u}_y+q_2{\overline u}_y^2},\quad R_{yz}={q_0\over1-q_1{\overline
u}_x+q_2{\overline u}_x^2}, \label{stressfitxy} \EN where $q_0$, $q_1$, and
$q_2$ are functions of ${\cal R}$ that can be determined by numerically fitting
$R_{xz}$ to ${\overline u}_y$ and $R_{yz}$ to $-{\overline u}_x$. It turns out
that in these two cases the three fit coefficients are the same to a good
approximation, in agreement with the results by Sulem \ea (1989). We have
therefore determined the best fit for the combined data set. The results are
shown in \Tab{Tfit} and \Fig{Fpstress_avero_comp}.

\begin{table}[t!]\caption{
Normalized fit coefficients $q_0$, $q_1$,
and $q_2$ for models with different values of ${\cal R}$.
}\vspace{12pt}\centerline{\begin{tabular}{c|cc|ccc}
Run & $k_{\rm f}/k_1$ & ${\cal R}$ & $q_0/(\nu k_{\rm f})^2$ &
$q_1\times(\nu k_{\rm f})$ & $q_2\times(\nu k_{\rm f})^2$ 
\\
\hline
1a &  6 & 0.82 & 0.31  & 0.72  & 0.38  \\ 
2a &  6 & 1    & 0.25  & 0.73  & 0.35  \\ 
3a &  6 & 1.2  & 0.35  & 0.62  & 0.28  \\ 
4a &  6 & 1.5  & 0.52  & 0.52  & 0.23  \\ 
5a &  6 & 2    & 0.81  & 0.41  & 0.19  \\ 
6a &  6 & 2.3  & 0.98  & 0.34  & 0.17  \\ 
7a &  6 & 3    & 1.19  & 0.18  & 0.07  \\ 
7b &  6 & 3    & 2.23  & 0.09  & 0.03  \\ 
8a &  6 & 5    & 4.11  & 0.043 & 0.012 \\ 
9a &  6 & 8    & 6.99  & 0.016 &-0.003 \\ 
9c & 14 & 8    & 7.20  & 0.009 & 0.005 \\ 
10a&  6 & 12   & 10.3  & 0.010 & 0.002    
\label{Tfit}\end{tabular}}\end{table}

\begin{figure}[h!]\epsfbox{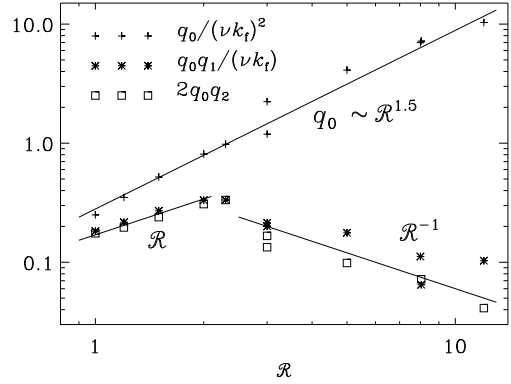}\caption[]{The ${\cal R}$
dependence of $q_0/(\nu k_{\rm f})^2$, $q_0q_1/(\nu k_{\rm f})$, and
$2q_0q_2$.}\label{Fpq}\end{figure}

Again, expanding \Eq{stressfitxy} we have
\EQ
R_{xz}\approx q_0-q_0q_1{\overline u}_y,\quad
R_{yz}\approx q_0+q_0q_1{\overline u}_x,
\label{stressfitxy_expand}
\EN
and therefore
\EQ
\alpha_{xzy}=-\alpha_{yzx}=q_0q_1.
\EN
It turns out that to a good approximation the values of
$q_0q_1/(\nu k_{\rm f})$ and $2q_0q_2$ agree with each other;
see \Fig{Fpq}. For ${\cal R}\leq2$, $q_0q_1/(\nu k_{\rm f})$ increases
with ${\cal R}=U_0/(\nu k_{\rm f})$. This implies that in dynamical
units $q_0q_1$ is constant: $q_0q_1/U_0\approx0.17$ for ${\cal R}\leq2$.
For larger values of ${\cal R}$, $q_0q_1/(\nu k_{\rm f})$ decreases like
${\cal R}^{-1}$. In dynamical units, $q_0q_1$ decreases quadratically:
$q_0q_1/U_0\approx0.6{\cal R}^{-2}$.

\epsfxsize=8cm

\epsfxsize=8cm

\begin{figure}[h!]\epsfbox{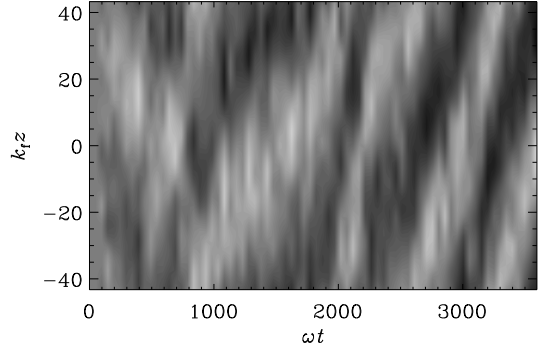}\caption[]{
Time-space diagram of ${\overline u}_x(z,t)$ for Run~7f. $k_{\rm f}/k_1=14$,
${\cal R}=3$. Positive values of ${\overline u}_x$ are shown in light and
negative values in dark. Note the migration in the positive $z$-direction.
}\label{Fpbutter_uu}\end{figure}

As we have already seen in \Fig{Fpuls}, the solutions change cha\-rac\-ter near
${\cal R}=3$. It turns out that in some cases with ${\cal R}\ge3$ the
solutions come in the form of travelling waves; see \Fig{Fpbutter_uu}.
For some values of ${\cal R}$, however, we also found standing waves
that alternated in time. A typical value for the period is $\omega T=600$.

\epsfxsize=8cm

\subsection{Magnetic fields}
\label{Smagnetic}

\begin{figure}[h!]\epsfbox{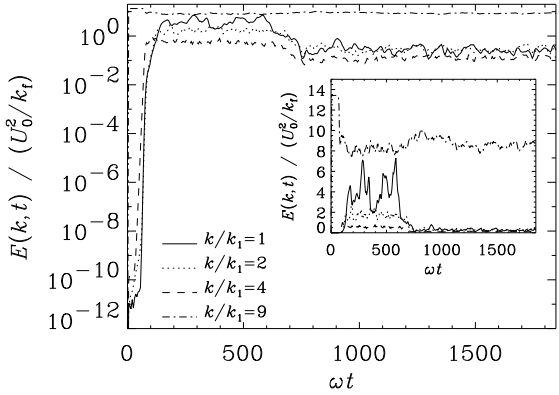}\caption[]{
Evolution of spectral energy in different modes in a lin-log plot. The inset shows
the same, but here the ordinate scale is linear. Run~7g.}
\label{Fpekt_akag4c}\end{figure}

We now discuss runs where the value of ${\cal R}_{\rm m}$ is large enough to
allow dynamo action. A good example is Run~7g where ${\cal R}_{\rm m}=12$. Here
the kinematic growth rate of the magnetic field is $\lambda=0.02\omega$.
The initial field has grown by a bit
more than five orders of magnitude before it reaches saturation, which is when
$\omega t\approx\ln10^5/0.02\approx600$. One clearly sees that at this time the
large scale flow becomes strongly suppressed; see \Fig{Fpekt_akag4c}, where we
have plotted the evolution of kinetic energies contained in various Fourier
modes. The normalized magnetic field strength, ${\cal B}$, saturates at a value
similar to the normalized velocity, ${\cal U}$. Looking at \Tab{T1}, it is
clear that for all runs with a magnetic field the large scale flow is strongly
suppressed, i.e.\ ${\cal U}_{\rm LS}/{\cal U}$ never exceeds the value $0.1
\dots 0.2$.

\epsfxsize=8cm

\epsfxsize=8cm\begin{figure}[h!]\epsfbox{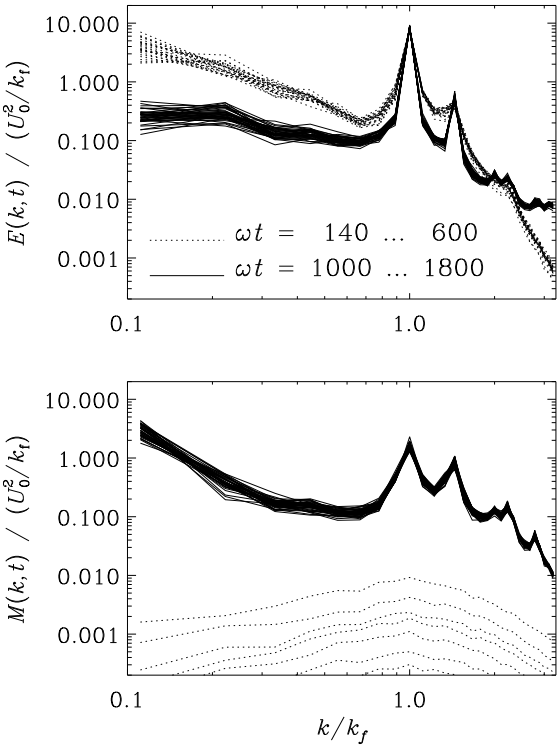}\caption[]{Power spectra
of kinetic (upper panel) and magnetic (lower panel) energies for Run~7g at different
times before saturation of the
large scale fields (dotted lines) and after saturation (solid lines). Note that
after saturation the inverse cascade-type behaviour in velocity is replaced by
an inverse cascade-type behaviour of the magnetic field.
}\label{Fpspec_dynb}\end{figure}

Prior to saturation of the magnetic field the kinetic energy spectrum looks
like that in \Fig{Fpspec_Aka3}, with peaks at both the forcing scale as well as
the largest scale of the system; see \Fig{Fpspec_dynb}. (The multiple peaks to the right of
the forcing wavenumber come from higher harmonics that are here more
pronounced than at higher values of ${\cal R}$; cf.\ \Fig{Fpspec_Aka3}.)
However, as the magnetic energy approaches
saturation, the kinetic energy becomes suppressed at large scales.
\FFig{Fpspec_dynb} shows that, in the saturated state, the kinetic energy
spectrum has lost its second peak at $k=k_1$. Instead, the magnetic energy
spectrum has now attained a shape similar to that of the kinetic energy before
saturation. In that sense, the inverse cascade-type behaviour in velocity is
now replaced by an inverse cascade-type behaviour of the magnetic field.

\epsfxsize=8cm\begin{figure}[h!]\epsfbox{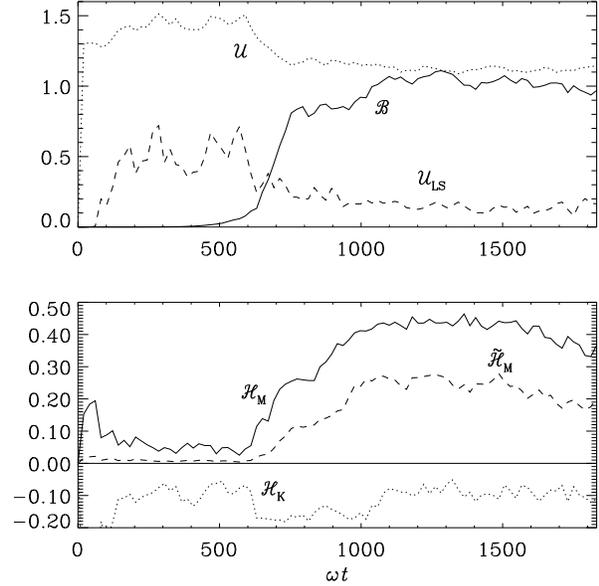}\caption[]{Evolution of
${\cal U}$, ${\cal U}_{\rm LS}$, and ${\cal B}$ (upper panel) and ${\cal
H}_{\rm K}$, ${\cal H}_{\rm M}$, and $\tilde{\cal H}_{\rm M}$ (lower panel).
Run~7g.}\label{Fphels}\end{figure}

In \Fig{Fphels} we have plotted the evolution of ${\cal U}$,
${\cal U}_{\rm LS}$, and ${\cal B}$ (upper panel) together with the relative
kinetic and magnetic helicities, ${\cal H}_{\rm K}$ and ${\cal H}_{\rm M}$,
as well as the nondimensional length scale related to
the magnetic helicity, $\tilde{\cal H}_{\rm M}$ (lower panel).
The kinetic helicity is always negative and temporarily enhanced around
the time when the magnetic field reaches saturation. We recall
that at each instant the forcing has zero helicity, and it is only due to the temporal
shift of the forcing pattern that helicity is introduced into the flow.

The magnetic field has positive helicity (\Fig{Fphels}). The opposite signs of
kinetic and magnetic helicities are in agreement with what is found for
helically forced turbulence; see Brandenburg (2001), where the forcing was
however chosen to have positive helicity. This led to positive helicity of the
flow, so all signs are reversed compared to the present case. Indeed, closer
inspection of the power spectrum of the magnetic helicity shows that at small
scales the signs of kinetic and magnetic helicities agree. However, because of
(approximate) helicity conservation the integrated helicity spectrum has to
vanish, which is why at large scales the sign of magnetic helicity is reversed.

\section{Conclusions}
\label{Sconcl}

The present calculations have verified the possibility of the formation of
large scale flows from small scale flows lacking parity invariance. These large
scale flows tend to be helical and of Beltrami type, as expected from the
nature of the anisotropic kinetic alpha (or AKA) effect. The power of the large
scale flows tends to be more strongly distributed over several Fourier modes.
This is in stark contrast to the magnetic inverse cascade or the dynamo alpha
effect which tends to select just one large scale Fourier mode -- at least in
the case of a fully periodic box; see Brandenburg (2001).

The resulting flows support the possibility of dynamo action once the magnetic
Reynolds number exceeds a value typical of non-helical dynamo action. This
critical value of the magnetic Reynolds number, based on the wavenumber of the
forcing, is around 10. For comparison, in the case of fully helical forcing,
the critical dynamo numbers are about ten times smaller; cf.\ Brandenburg
(2001), where the magnetic Reynolds numbers based on the forcing scale need to
be divided by $2\pi$ to be compatible with the present work. However, once the
magnetic field reaches saturation, it begins to suppress the large scale flow.

In the present work we have adopted the forcing function of Frisch \ea (1987).
This has the advantage that we were able to make contact with earlier studies
where the AKA effect was also considered numerically. This flow was constructed
on purely mathematical grounds in order to demonstrate the very existence of
the AKA effect. Other more realistic types of forcing have been proposed by
Levina \ea (2000). It is important to extend the present studies to these types
of forcing in order to see whether the AKA effect works for broader classes of
forcing. However, our present results suggest that the
anisotropic kinetic alpha effect should not play a role in astrophysics where
the value of ${\cal R}$ is in general large and magnetic fields are usually
present. It should be emphasized that in some of the best cases where a large
scale velocity pattern emerged (e.g.\ Run~7f), the large scale flow pattern is
never really as pronounced as the large scale magnetic fields that are
produced by helical turbulence (cf.\ Brandenburg 2001).

Finally, we should mention that our results do by no means address the question
of whether or not large scale vortices can form astrophysical bodies such as
giant planets or accretion discs. Such vortices are long-lived, quasi-stable
formations, possibly belonging to the class of solutions studied by Goodman \ea
(1987). It it possible that they are formed simply as a matter of suitable
initial conditions, but they were also found in simulations of Hawley (1987).
This type of solution would be essentially nonlinear, in contrast to
solutions of the mean-field equations with AKA effect which are possible
already in linear approximation.

\begin{acknowledgements}
We thank Wolfgang Dobler and Galina Levina for making useful suggestions to the manuscript.
Use of the PPARC supported supercomputers in St Andrews and Leicester (UKAFF)
is acknowledged.
\end{acknowledgements}

\bigskip\noindent{\it
$ $Id: paper.tex,v 1.66 2001/06/15 13:31:19 brandenb Exp $ $}

\end{document}